\documentclass[aps,prl,showpacs,twocolumn,preprintnumbers,amsmath,amsfonts,bm]{revtex4-1}
\usepackage{graphicx}
\usepackage{subfigure}
\usepackage{dcolumn}

\begin{document}

\title{Emergent Criticality Through Adaptive Information Processing in
  Boolean Networks}

\author{Alireza Goudarzi$^{1}$, Christof Teuscher$^{1}$, Natali
  Gulbahce$^{2}$ and Thimo Rohlf$^{3,4}$}

\affiliation{
$^1$Portland State University, 1900 SW $4^{th}$ Ave, Portland, OR 97206 USA\\
$^2$University of California, San Francisco, 1700 $4^{th}$, San Francisco, CA 94158 USA\\
$^3$Interdisciplinary Center for Bioinformatics, University Leipzig, Haertelstr. 16-18, D-04107 Leipzig, Germany\\
$^4$Max-Planck-Institute for Mathematics in the Sciences, Inselstr. 22, D-04103 Leipzig, Germany.\\
}

\date{\today}

\begin{abstract}
  We study information processing in populations of Boolean networks
  with evolving connectivity and systematically explore the interplay
  between the learning capability, robustness, the network topology,
  and the task complexity. We solve a long-standing open question and
  find computationally that, for large system sizes $N$, adaptive
  information processing drives the networks to a critical
  connectivity $K_{c}=2$. For finite size networks, the connectivity
  approaches the critical value with a power-law of the system size
  $N$.  We show that network learning and generalization are optimized
  near criticality, given task complexity and the amount of
  information provided surpass threshold values.  Both random and
  evolved networks exhibit maximal topological diversity near
  $K_{c}$. We hypothesize that this diversity supports efficient
  exploration and robustness of solutions. Also reflected in our
  observation is that the variance of the fitness values is maximal in
  critical network populations.  Finally, we discuss implications of
  our results for determining the optimal topology of adaptive
  dynamical networks that solve computational tasks.
\end{abstract}


\maketitle

In 1948, Alan Turing proposed several unorganized machines made up
from randomly interconnected two-input NAND logic gates
\cite{turing69} as a biologically plausible model for computing. He
also proposed to train such networks by means of a ``genetical or
evolutionary search.'' Much later, random Boolean networks (RBN) were
introduced as simplified models of gene regulation
\cite{Kauffman69,Kauffman93}, focusing on a system-wide perspective
rather than on the often unknown details of regulatory interactions
\cite{Bornholdt2001}. In the thermodynamic limit, these disordered
dynamical systems exhibit a dynamical order-disorder transition at a
sparse critical connectivity $K_c$ \cite{DerridaP86}.  For a finite
system size $N$, the dynamics of RBNs converge to periodic attractors
after a finite number of updates. At $K_c$, the phase space structure
in terms of attractor periods \cite{AlbertBaraBoolper00}, the number
of different attractors \cite{SamuelsonTroein03} and the distribution
of basins of attraction \cite{Bastola98} is complex, showing many
properties reminiscent of biological networks \cite{Kauffman93}.  In
cellular automata (CA), complex computation has been hypothesized to
occur where the rules show complex dynamics at ``the edge of chaos''
\cite{langton1990,packard1988}.  This claim was refuted in
\cite{mitchell93}. However, the argument in \cite{mitchell93} rests on symmetric spaces
in the CA lattice and rule space. These results therefore do not apply
to RBN.  Phase transition in information dynamics was studied in
\cite{lizier2008}.  State-topology coevolution in RBNs was studied by
\cite{Kauffman:1991p959,BornholRohlf00,liu2006} and it was shown that
networks evolved toward a critical connectivity $K_c=2$. This letter
presents the first study to link complex dynamics, topology, and task
solving in an open RBN.

In \cite{carnevali87a,carnevali87b,patarnello89,broeck90} simulated
annealing (SA) and genetic algorithms (GAs) were used to train
feedforward RBNs and to study the thermodynamics of learning.  For a
given task with predefined input-output mappings, only a fraction of
the input space is required to train networks that
generalize perfectly on all input patterns. This fraction depends on
the network size and the task complexity. Moreover, the more inputs a
task has, the smaller the training set needs to be to obtain full
generalization.  In this context, {\em learning} refers to correctly
solving the task for the training samples while {\em generalization}
refers to correctly solving the task for novel inputs. We use {\em
  adaptation} to refer to the phase where networks have to adapt to
ongoing mutations (i.e., noise and fluctuations), but have already
learned the input-output mapping. In this Letter, we study adaptive
information processing in populations of Boolean networks with an
evolving topology. Rewiring of connections and mutations of the
functions occur at random, without bias toward particular topologies
(e.g., feedforward).  We systematically explore the interplay between
the learning capability, the network topology, the system size $N$,
the training sample $T$, and the complexity of the computational task.

First, let us define the dynamics of RBNs. A RBN is a discrete
dynamical system composed of $N$ automata.  Each automaton is a
Boolean variable with two possible states: $\{0,1\}$, and the dynamics
is such that ${\bf F}:\{0,1\}^N\mapsto \{0,1\}^N, $ where ${\bf
  F}=(f_1,...,f_i,...,f_N)$, and each $f_i$ is represented by a
look-up table of $K_i$ inputs randomly chosen from the set of $N$
automata. Initially, $K_i$ neighbors and a look-table are assigned to
each automaton at random. For practical reasons we restrict the
maximum $K_i$ to $8$. An automaton state $ \sigma_i^t \in \{0,1\}$ is
updated using its corresponding Boolean function, $\sigma_i^{t+1} =
f_i(\sigma_{i_1}^t,\sigma_{i_2}^t, ... ,\sigma_{i_{K_i}}^t)$.

The automata are updated synchronously using their corresponding
Boolean functions.  For the purpose of solving computational tasks, we
define $I$ inputs and $O$ outputs. The inputs of the computational
task are randomly connected to an arbitrary number of automata. The
connections from the inputs to the automata are subject to rewiring
and are counted to determine the average network connectivity $\langle
K \rangle$. The outputs are read from a randomly chosen but fixed set
of $O$ automata. All automata states are initialized to ``$0$'' for
each input pattern before simulating the network.

{\em Methodology.}--- We evolve the networks by means of a traditional
genetic algorithm (GA) to solve three computational tasks of varying
difficulty, each of which defined on a $3$-bit input: full-adder (FA),
even-odd (EO), and the cellular automata rule $85$ (R85)
\cite{wolfram83}. The FA task receives two binary inputs $A$, $B$, an
input carry bit $C_{in}$, and outputs the binary sum of the three
inputs $S=A+B+C_{in}$ on the $2$-bit output and the carry bit
$C_{out}$. The EO task outputs a $1$ if there is an odd number of $1$s
in the input (independent of the order), a $0$ otherwise. R85 is
defined for three binary inputs $A$, $B$, and $C$, and outputs the
negation of $C$. The output for R85 task therefore only depends on
one input bit. The EO task represents the most difficult task,
followed by the FA and R85 task. Task difficulty is the complexity of
information integration needed in the input to determine the
output. This can be measured through information-theoretical
decomposability of a task. We can represent the task itself as the
contingency table of its inputs and outputs. Different decomposition
models of the task are the different ways that we can calculate the
marginal probabilities from the original contingency table
\cite{zwick2004}. We calculate a weighted sum of the vector of the
information content of all possible decomposition models of a task
$F$. This can be summarized in: ${{decomposition}_{\text  F}=\sum_{\text m\in
  Models_F}{\text w_m Inf_m}}$, where $Models_F$ is the set of all
decomposition models of $F$, and the weight $w_m$ of a model is
proportional to its degrees of freedom. The information content of a
model is calculated using $Inf_m=1-\frac{\text H_m - \text H_F}{\text H_{ind} - \text
  H_F}$. Here, $H_m$ is the entropy of the model, $H_F$ is the entropy
of $F$, and $H_{ind}$ is the entropy of the independence model (all
input and output variables are assumed independent). Higher values for
${decomposition}_F$ mean that the task is more decomposable and
therefore less difficult.

The genetic algorithm we use is mutation-based only, i.e., no
cross-over operation is applied.  For all experiments we ran a
population of $30$ networks with initial connectivity $\langle
K_{in}\rangle=1$ and a mutation rate of $0.8$.  Each mutation is decomposed
into $1+\alpha$ steps repeated with probability
$p(\alpha)=0.5^{\alpha+1}$, where $\alpha \ge 0$. Each step involves
flipping a random location of the look-up table of a random automata
combined with adding or deleting one link. Each population is run for
$30,000$ generations. We repeat each evolutionary run $30$ times and
average the results. In each generation and for each tested input
configuration, the RBN is run for a convergence time $t\propto N$ updates. {Afterward, we run the network for an
  additional $t\propto N$ updates to record the activity of the output
  nodes. If the activity of an output node is ``1'' for at least half
  of the $t$ time steps, we interpret the output as a ``1'', and as
  ``0'' otherwise.} For an evolutionary run of training size $T$, the
training sample set $M$ is randomly chosen at each generation without
replacement from the $2^3$ possible input patterns. During each
generation, the fitness of each individual is determined by $f=1-E_M$,
where $E_M$ is the normalized average error over the $T$ random
training samples: $E_M=\frac{1}{T}\sum_{i\in M}{\sum_{j\in O}{(a_{ij}
    - o_{ij})^2}}$.  $a_{ij}$ is the value of the output automata $j$
for the input pattern $i$, and $o_{ij}$ is the correct value of the
same bit for the corresponding task. The generalization score is
calculated using the same equation with $M$ including all $2^3$ inputs
rather than a random sample. Finally, selection is applied to the
population as a deterministic tournament. Two individuals are picked
randomly from the old population and their fitness values are
compared. The better individual is mutated and inserted into the new
population, the worse individual is discarded. We repeat the process
until we have $30$ new individuals in the new population.
 
\begin{figure}
\centering 
\includegraphics[width=2.9in]{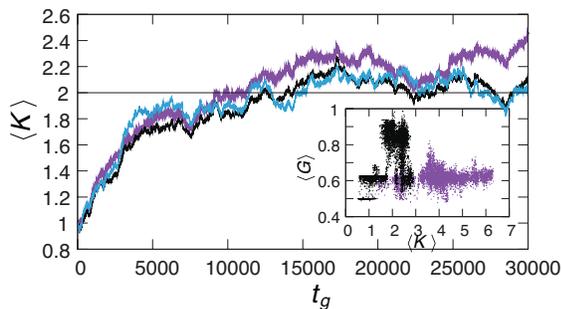}
\vspace{-5mm}
\caption{Convergence of the average network connectivity as a function
  of the GA generations $t_g$. FA task with $T=4$ and $N=100$. The
  curves are averaged over $30$ evolutionary runs (red), only the $22$
  best (green), and the $15$ best (light blue) populations,
  respectively. Inset: scatter plot correlating average $\langle K
  \rangle(t_g)$ and average generalization $\langle G \rangle(t_g)$ of
  a successful population (black) and a suboptimal population
  (purple).}
\label{fig:tempdyn}
\end{figure}

\begin{figure}
\centering
\includegraphics[width=3.5in]{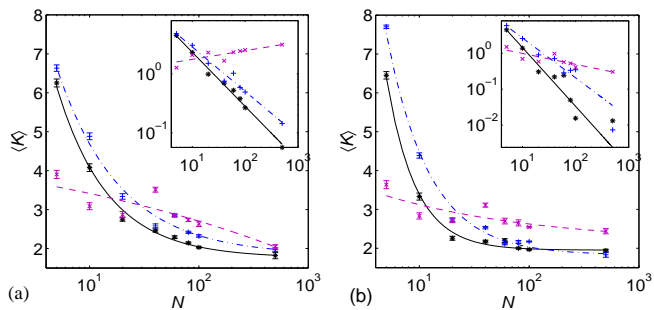}
\vspace{-7mm}
\caption{Finite size scaling of $\langle K \rangle$ as a function of
  $N$ for the three tasks, EO (black), FA (blue), R85 (magenta), and
  the training sample size $T=4$ (a) and $T=8$ (b).  Points represent
  the data of the evolved networks, lines represent the fits. The
  finite size scaling for $\langle K \rangle$ shows that it scales
  with a power-law as a function of the system size $N$. The dashed
  lines represent the power-law fit of the form $a*x^b+c$. We favor
  the data for larger $N$ by weighting the data according to
  $N/N_{max}$, where $N_{max}=500$. {The insets show
    $K_c-c$ as a function of $N$ on a log-log scale.}}
\label{fig:scaling}
\end{figure}

{\em Results.}--- We observe a convergence of $\langle K \rangle$
close to the critical value $K_c = 2$ for large system sizes $N$ and
training sample sizes larger or equal to $T=4$. For $T=8$, populations
always evolve close to criticality for moderate $N$. For
smaller $T$, the average over all evolutionary runs is found at
slightly higher values of $\langle K \rangle$
(Fig.~\ref{fig:tempdyn}). If the average is taken only over the {\em
  best individuals}, however, $\langle K \rangle$ values close to
$K_c$ are recovered. This observation can be explained from the fact
that for $T<8$, due to the limited information provided for learning,
some populations cannot escape local optima, and hence do not reach
maximum fitness.  Sub-optimal network populations show a large scatter
in $\langle K \rangle$ values in the evolutionary steady state, while
those with high fitness scores cluster around $K_c=2$
(Fig.~\ref{fig:tempdyn}, inset). For the simple R85 task we do not
observe any convergence to $K_c = 2$, independent of the training
samples. For the other tasks, the finite size scaling of $\langle K
\rangle$ (Fig.~\ref{fig:scaling}) exhibits convergence towards $K_c$
with a power-law as a function of the system size $N$. For $T=8$, the
exponent $b$ of the power-law for the three tasks EO, FA, and R85 is
$-1.63$, $-1.11$, and $-0.30$ respectively (Fig.~\ref{fig:scaling}).
Altogether, these results suggest that the amount of information
provided by the input training sample helps to drive the network to a
critical connectivity.

{Interestingly, the population dynamics in our model
  follow Fisher's fundamental theorem of natural selection, which
  attributes the rate of increase in the mean fitness to the increased
  fitness variance in the population \cite{edwards1994}. It has been
  shown in GAs that the diversity maximization
  \cite{Bedau94bifurcationstructure} makes more configurations of the
  search space accessible to the genetic search to find optimal
  solutions \cite{carnevali87a,carnevali87b,patarnello89}.}

\begin{figure}
\centering 
\includegraphics[width=3.2in]{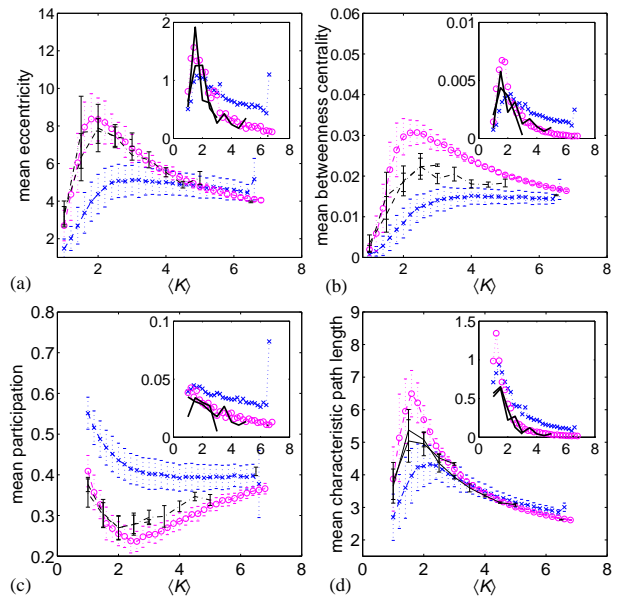}
\vspace{-3mm}
\caption{Near $K_c$, the topology of the network shows maximal
  variance. The insets show the standard deviation of the topological
  measures for initial ER networks (magenta), the evolved networks
  (black), and the XRG (blue). The solid lines represent the
  topological measures in random networks. The dotted line represents
  the same measure in RBNs.}
\label{fig:topology}
\end{figure}

\begin{figure}
\centering  
\includegraphics[width=2.6in]{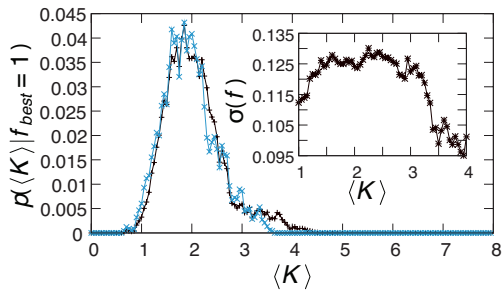}
\vspace{-3mm}
\caption{The conditional probability that evolving populations, where
  the best mutant reaches maximum fitness (i.e., $f_{best} = f_{max} =
  1$), have average connectivity $\langle K \rangle$ shows a sharp
  peak near $K_c$ (black curve), the same is found for maximum
  generalization (light blue). Inset: diversity of evolving populations,
  quantified in terms of the standard deviation $\sigma(f)$ of fitness
  distributions, has a maximum near $K_c=2$. All data sampled over the
  best $22$ out of $30$ populations for full-adder task with $T=4$ and
  $N=100$.}
\label{fig:bestmutants_of_k} 
\end{figure}

{Indeed, we find that the standard deviation of the fitness values in
the populations has a local maximum near $K_c$
(Fig.~\ref{fig:bestmutants_of_k}, inset), with a sharp decay toward
larger $\langle K \rangle$, indicative of maximum diversity near criticality.}
 Evidently, this diversity helps to
maintain a high fitness population in the face of continuous mutations
with a fairly high rate ($0.8$ in our study). While the average
fitness can be lower (and often is), compared to less diverse
populations, the probability to find and maintain high fitness
solutions is strongly increased. Indeed, we find that populations
where the best mutant has maximum fitness ($f=1$) sharply peak near
$K_c$ (Fig.~\ref{fig:bestmutants_of_k}), as well as populations where
the best mutant reaches perfect generalization. {To find a possible
source of fitness diversity, we determined several topological
measures of the networks \cite{rubinov2010}: the eccentricity (maximum shortest path between a vertex $v$ and any other vertex in a graph), the
betweenness centrality (the average fraction of shortest paths between all vertices in a graph that passes through a vertex $v$), the participation, and the characteristic path length.
These measures were calculated for Erd{\"o}s-R{\'e}nyi (ER), eXponential Random Graphs (XRG), as well
as for the evolved networks (Fig.~\ref{fig:topology}).}  In fact, we find
that the graph-theoretical measures have maximal variance near
$K_c=2$. Similarly, other authors have shown that dynamical diversity
is maximized near $K_c$, too \cite{Nykter2008}. Our results suggest
that evolving RBN can indeed exploit this diversity to optimize
learning.
 
In addition, we find that during the learning process of the networks,
the in-degree distribution changes from a Poissonian to an exponential
distribution. In particular, we observe that the topological
properties of the networks reach a compromise between ER graphs and
the XRG. The same observation was made in input- and output-less RBNs
that were driven to criticality by using a local rewiring rule
\cite{BornholRohlf00}.  This significant topology change is related to
diversity (entropy) maximization during the learning phase
\cite{RevModPhys.80.1275}. However, this is beyond the scope of this
paper and will be discussed in a separate publication.

{Finally, we measured the damage spreading in the
  evolved RBNs \cite{DerridaP86} to determine their dynamical
  regime. The damage spreading $d_{t+1}$ is measured by changing the
  state of a randomly selected node in two identical networks. The two
  networks are simulated for a single time step and the damage
  spreading $\bar d$ is then calculated by averaging the ratio
  $\frac{d_{t+1}}{d_t}$ over many trails with random initial network
  configurations.  One observes that for critical networks $\bar d=1$, for
  supercritical networks $\bar d>1$, and for subcritical networks
  $\bar d<1$. We see that for networks with a high fitness, $\bar
  d$ peaks around $1$ for all $N$.}

{\em Discussion.}--- We investigated the learning and generalization
capabilities in RBNs and showed that they evolve toward a critical
connectivity of $K_c \approx 2$ for large networks and large input
sample sizes.  For finite size networks, the connectivity approaches
the critical value with a power-law of the system size $N$.  We showed
that network learning and generalization are optimized near
criticality, given task complexity and the amount of information
provided surpass threshold values.  Furthermore, critical RBN
populations exhibit the largest diversity (variance) in fitness
values, which supports learning and robustness of solutions under
continuous mutations.  By considering graph-theoretical measures, we
determined that $K_c$ corresponds to a region in network ensemble
space where the topological diversity is maximized, which may explain
the observed diversity in critical populations.
 
Interestingly, we observe that RBN populations that are optimal with
respect to learning and generalization tend to show average
connectivity values slightly below $\langle K\rangle = 2$. This may be
related to previous results indicating that $K_c < 2$ in finite size
RBN \cite{RohlfGulbahceTeuscher2007}.

Examination of the attractors of the final population confirms that
the computation happens as partitioning of the state-space into
disjoint attractors \cite{krawitz2007}.  During the evolution, the
attractor landscape changes so that there are enough attractors to
properly process the inputs. The entire task is encoded as a hyper
cycle (i.e., a set of mutually reachable attractors) in the network
dynamics. The input combinations play the role of a multi-valued
switch that pushes the dynamics out of one attractor into the next
along the hyper cycle. Emergence of the large attractor basins make
the computation highly robust to perturbations in the node state while
maintaining sensitivity to input signals. {All
  networks in our final population converge to fixed-point or cyclic
  attractors.}

To summarize, we solved a long-standing question and showed that
learning of classification tasks and adaptation can drive RBNs to the
``edge of chaos'' \cite{Kauffman93}, where high-diversity populations
are maintained and on-going adaptation and robustness are optimized.
Our study may have important implications for determining the optimal
topology of a much larger class of complex dynamical networks where
adaptive information processing needs to be achieved efficiently,
robustly, and with limited connectivity (i.e., resources). This has
applications, e.g., in the area of neural networks, complex networks,
and more specifically in the area of emerging molecular and nanoscale
networks and computing devices, which are expected to be built in a
bottom-up way from vast numbers of simple, densely arranged components
that exhibit high failure rates, are relatively slow, and connected in
an unstructured way.

{\em Acknowledgments.}  This work was partly funded by NSF grant \#
1028120. The first author is grateful for the fruitful discussions
with Guy Feldman and Lukas Svec.

\end{document}